\documentclass[a4paper,10pt]{article}

\usepackage{amsmath,epsfig,amsfonts,  bezier,amstext,amsthm,makeidx,indentfirst,ifthen, minitoc,amssymb,amsbsy,color}
\usepackage{multirow}
\usepackage[T1]{fontenc}                      
\usepackage[english]{babel}
\usepackage{graphicx,xr}
\usepackage{float}
\usepackage{mathrsfs}
\usepackage{rotating}
\usepackage{lscape}
\usepackage{makecell}
\usepackage{threeparttable}
\usepackage[normalem]{ulem}
\usepackage{epstopdf} 
\usepackage{setspace}
\usepackage{url}
\usepackage{overpic}

\usepackage{rotating}
\usepackage[section]{placeins}
\newcolumntype{C}[1]{>{\centering\arraybackslash}p{#1}}
\usepackage{comment}
\usepackage{csquotes}
\usepackage{caption}
\usepackage{subcaption}
\usepackage{comment}
\usepackage[usenames,dvipsnames,svgnames,table]{xcolor}
\usepackage{tabularx,colortbl}
\usepackage{multicol}
\usepackage{etoolbox}
\newcommand{\be}{\begin{equation}}
\newcommand{\ee}{\end{equation}}

\newcommand{\del}[2]{\frac{\partial#1}{\partial#2}}

\newcommand{\Ubold}{\textbf{u}}

\definecolor{amethyst}{rgb}{0.6, 0.4, 0.8}

\usepackage{array}
\newcommand*{\vertbar}{\rule[-1ex]{0.5pt}{2.5ex}}

\begin{document}
{\vskip -1cm}

\title{\LARGE{\textbf{Nonlinear parametric models of viscoelastic fluid flows} }}

\author{{Cassio M. Oishi}\thanks{\,e-mail:
cassio.oishi@unesp.br} \\
{\normalsize Departamento de Matem\'atica e Computa\c c\~ao, Faculdade de Ci\^encias e Tecnologia, }\\
{\normalsize São Paulo State University, Presidente Prudente, Brazil} \\ [2ex]
{Alan A. Kaptanoglu}\thanks{\,e-mail: akaptano@umd.edu}\\
{\normalsize IREAP, University of Maryland, College Park, EUA}\\
{\normalsize Department of Mechanical Engineering,}\\
{\normalsize University of Washington, Seattle, EUA}\\[2ex]
{J. Nathan Kutz}\thanks{\,e-mail: kutz@uw.edu}\\
{\normalsize Department of Applied Mathematics,}\\
{\normalsize University of Washington, Seattle, EUA}\\[2ex]
{Steven L. Brunton}\thanks{\,e-mail: sbrunton@uw.edu}\\
{\normalsize Department of Mechanical Engineering,}\\
{\normalsize University of Washington, Seattle, EUA}\\
}

\date{}
\maketitle

\vskip -1cm
\begin{abstract}
 Reduced-order models have been widely adopted in fluid mechanics, particularly in the context of Newtonian fluid flows. These models offer the ability to predict complex dynamics, such as instabilities and oscillations, at a considerably reduced computational cost. 
 In contrast, the reduced-order modeling of non-Newtonian viscoelastic fluid flows remains relatively unexplored.  
 This work leverages the {\em sparse identification of nonlinear dynamics} (SINDy) algorithm to develop interpretable reduced-order models for viscoelastic flows\footnote{Video research abstract: \url{https://www.youtube.com/watch?v=pBAmBZP5Sp8}}. 
 In particular, we explore a benchmark oscillatory viscoelastic flow on the four-roll mill geometry using the classical Oldroyd-B fluid. 
 This flow exemplifies many canonical challenges associated with non-Newtonian flows, including transitions, asymmetries, instabilities, and bifurcations arising from the interplay of viscous and elastic forces, all of which require expensive computations in order to resolve the fast timescales and long transients characteristic of such flows.  
 First, we demonstrate the effectiveness of our data-driven surrogate model to predict the transient evolution and accurately reconstruct the spatial flow field for fixed flow parameters.  We then develop a fully parametric, nonlinear model capable of capturing the dynamic variations as a function of the Weissenberg number. While the training data is predominantly concentrated on a limit cycle regime for moderate $Wi$, we show that the parameterized model can be used to extrapolate, accurately predicting the dominant dynamics in the case of high Weissenberg numbers. 
 The proposed methodology represents an initial step in the field of reduced-order modeling for viscoelastic flows with the potential to be further refined and enhanced for the design, optimization, and control of a wide range of non-Newtonian fluid flows using modern machine learning and reduced-order modeling techniques.   

\end{abstract}

\textit{Keywords: Viscoelastic fluids, Computational fluid dynamics, Data-driven models, Sparse identification of nonlinear dynamics, Reduced-order models, Machine learning}

\newpage

\section{Introduction}
\label{intro}

Viscoelastic fluids are an important class of non-Newtonian materials that exhibit both viscous (liquid-like) and elastic (solid-like) properties. Due to their importance in a wide range of applications, computational methods have been widely adopted to solve viscoelastic fluid flows, leading to new insights in non-Newtonian mechanics. 
Despite advances in scientific computing for modeling and simulating viscoelastic fluid flows~\cite{Alves}, many problems remain computationally challenging, such as resolving viscoelastic instabilities and elastic turbulence~\cite{PhysRevFluids.7.080701}. 
Therefore, it is essential to develop enhanced numerical schemes for non-Newtonian fluid mechanics based on machine learning algorithms and data-driven strategies.

Reduced-order models (ROMs) have been successfully developed and applied to a wide range of Newtonian fluid flows~\cite{Noack2011book,noack2011galerkin,Carlberg2013jcp,Kaiser2014jfm,quarteroni2014reduced,Benner2015siamreview,Carlberg2015siamjsc,peherstorfer2016data,Carlberg2017jcp,Rowley2017arfm,qian2020lift,benner2020operator,kramer2021stability,Brunton2022book}.  
The goal is to develop a low-dimensional surrogate model that captures the dominant coherent behavior of a fluid flow at a fraction of the computational cost of a high-fidelity simulation.  
Classical approaches typically involve Galerkin projection of the governing Navier-Stokes equations onto a low-dimensional data-driven basis obtained via {\em proper orthogonal decomposition} (POD)~\cite{Noack2011book,noack2011galerkin,Benner2015siamreview}.  
The POD procedure is data-driven, providing a generalization of the Fourier transform that is tailored to a particular flow of interest.  
The Galerkin projection procedure results in a low-dimensional set of nonlinear ordinary differential equations for the amplitudes of these orthogonal modes, and these differential equations may be useful for efficient prediction, estimation, and control.  
{Although the POD step is data-driven, the Galerkin step is intrusive and often unstable~\cite{carlberg2017galerkin}, requiring a flexible working code to simulate the flow, and the nonlinearity typically becomes quite complex except for incompressible flows.}  
There are non-intrusive extensions based on operator inference~\cite{peherstorfer2016data,qian2020lift,benner2020operator,kramer2021stability} and several  extensions to exotic physics, including compressible flows~\cite{rowley2004model} and plasma physics~\cite{Kaptanoglu2021pre}.  
However, there is an increasing trend to replace the Galerkin projection step entirely with machine learning approaches~\cite{Brunton2020arfm,Duraisamy2019arfm,parish2020time,regazzoni2019machine}.

The sparse identification of nonlinear dynamics (SINDy)~\cite{Brunton} has been particularly useful for learning accurate and efficient dynamical systems models of complex fluid flows entirely from data.  
SINDy represents the differential equation governing POD mode amplitudes as a sparse combination from a library of candidate functions that might describe the dynamics; for fluid flows, this library is often chosen to be polynomial~\cite{Loiseau2017jfm}. 
SINDy has been widely applied to learn data-driven nonlinear models across a range of application domains, including fluid~\cite{Loiseau2017jfm,Loiseau2018jfm,loiseau2020data,guan2020sparse,deng2021galerkin,Callaham2022scienceadvances,Callaham} and plasma dynamics~\cite{Dam2017pf,Kaptanoglu2021pre}, turbulence closures~\cite{beetham2020formulating,beetham2021sparse,schmelzer2020discovery}, nonlinear optics~\cite{Sorokina2016oe}, and numerical integration schemes~\cite{Thaler2019jcp}.  
Similarly, several extensions have been introduced, including to identify partial differential equations~\cite{Rudy2017sciadv,Schaeffer2017prsa}, {tensor formulations~\cite{Gelss2019mindy}, control theory~\cite{Kaiser2018prsa}, integral and weak formulations~\cite{Schaeffer2017pre,Reinbold2020pre,gurevich2019robust,alves2020data,reinbold2021robust}, and systems with stochastic dynamics~\cite{boninsegna2018sparse,callaham2021nonlinear}.}   

In this work, we demonstrate the broad applicability of SINDy-based reduced-order modeling techniques for viscoelastic flows (see Fig. 1).  In particular, we develop a viscoelastic POD to identify coherent structures which can be well characterized by SINDy.  This work builds upon the limited ROM efforts in non-Newtonian flows which includes a POD-Galerkin reduced-order method for a viscoelastic model~\cite{WANG2020104747} and corresponding stabilized version of this method~\cite{Chetry}.  In both works, the number of POD modes required is quite high (e.g. $r > 30$). 
The current work shows that with partial knowledge of the physics, it is possible to develop accurate models with significantly fewer modes. 
Recently a similar approach has been successfully applied to study electroconvection~\cite{GUAN} and magnetohydrodynamics~\cite{Kaptanoglu2021pre}. 

From numerical measurement data of oscillatory flows on the four-roll mill geometry using the Oldroyd-B fluid, we first obtain a low-dimensioanl linear subspace by POD in order to extract dominant coherent structures in the flow. The SINDy algorithm is then applied to identify the dynamical system for flows characterized by a fixed set of parameters. We find a sparse reduced-order model that produces efficient and provably bounded forecasts of the solution of the complex dynamics, as well as accurately reconstructs the flow fields of the viscoelastic stress tensor on the governing equations. The elaborated methodology is then shown to be useful for learning parametric models that can capture variations in the Weissenberg number, an important non-dimensional parameter widely used to describe the elastic effects on viscoelastic flows~\cite{Alves,THOMPSON2021104550}.  

\begin{figure}[t!]
    \centering
    \includegraphics[width=\textwidth]{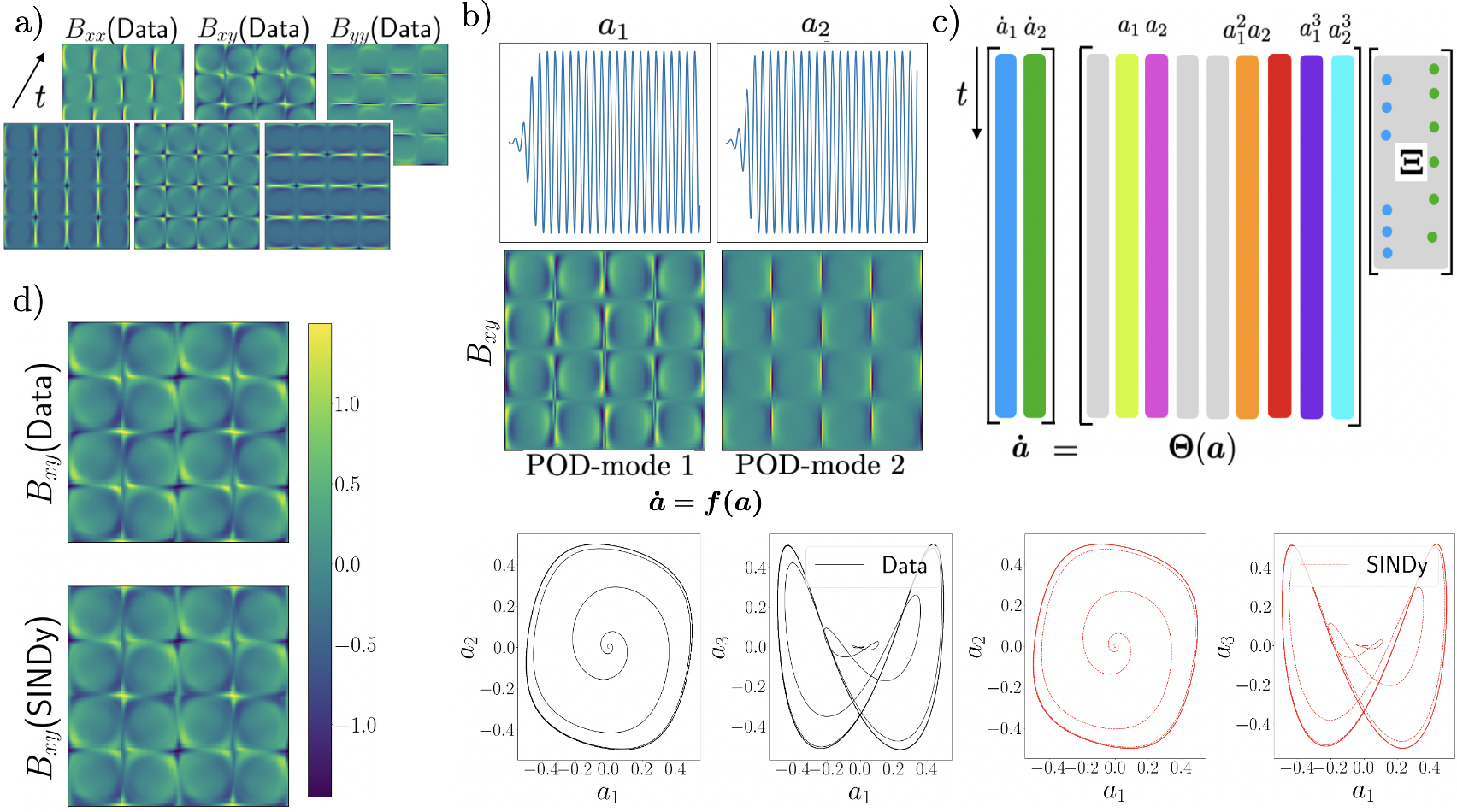}
    \caption{Summary of the sparse identification framework for a viscoelastic fluid flow: a) Construction of a data matrix using data from either simulations or experiments, b) Application of the viscoelastic proper orthogonal decomposition, c) Discovery data-driven using SINDy algorithm, d) Reconstruction of the flow field.}
    \label{illus_frm}
\end{figure}

\section{Governing equations and the four-roll mill flow}

In non-dimensional form, 
the mass and momentum equations combined with a viscoelastic constitutive equation for the Oldroyd-B fluid is given by~\cite{MARTINS2015653,Alves}

\begin{subequations}
\begin{align}
& \nabla \cdot \Ubold = 0, \label{eqAdim12}\\
& { \frac{\partial \Ubold}{\partial t} + \Ubold \cdot \nabla \Ubold = - \nabla p  + \frac{\beta}{Re}\nabla^2\Ubold + \frac{1}{Re}\nabla\cdot {\pmb{\tau}} + \bf{f}, } \label{eqAdim1}\\
& \del{\textbf{C}}{t} + (\Ubold\cdot\nabla)\textbf{C} = (\nabla\Ubold)\textbf{C} + \textbf{C}(\nabla\Ubold)^T  - \frac{1}{Wi}(\textbf{C}-\textbf{I}), \\
&   \pmb{\tau} = \frac{(1-\beta)}{Wi} (\textbf{C} - \boldsymbol{I}),
\label{eqAdim3}
\end{align}
\end{subequations}
where $\mathbf{u}$ is the fluid velocity, $p$ is the pressure,  $\textbf{C}$ is the conformation tensor, $\boldsymbol{\tau}$ the extra-stress tensor, and $\bf{f}$ is an external force term. The non-dimensional numbers in these equations are the Reynolds number ($Re$), the Weissenberg number ($Wi$), and the viscosity ratio ($\beta \in (0, 1]$) defined respectively as
$$
	{Re = \frac{\rho U L}{(\eta_s+\eta_p)}, \hspace{15pt} Wi = \frac{\lambda_p U}{L}, \hspace{15pt} \beta = \frac{\eta_s}{(\eta_s+\eta_p)}},
 $$
where $\rho$ is the fluid density, and $U$ and $L$ are the characteristic velocity and length scales, respectively. In addition, the fluid properties are described using the solvent viscosity, $\eta_s$, the polymeric viscosity, $\eta_p$, and the relaxation time $\lambda_p$. While the solvent viscosity is related to the Newtonian viscosity, the polymeric viscosity is directly related to the presence of the dissolved polymer. Therefore, the non-dimensional numbers $Re$ and $\beta$ can be defined using the total viscosity $\eta_0 = \eta_s + \eta_p$.

We will demonstrate the methods developed in this work on the canonical four-roll mill problem. 
This benchmark has been applied to investigate viscoelastic instabilities~\cite{GUTIERREZCASTILLO201948} and has led to new insights into elastic turbulence~\cite{dzanic_from_sauret_2022}. 
The geometry is given by a regular array of cylinders that are rotating, driving a rich sequence of behaviors and bifurcations.  
The presence of stagnation points between the rollers can cause high stretching of the polymer stress~\cite{doi:10.1126/science.276.5321.2016}, which is of particular interest for computation and experiment~\cite{PhysRevLett.109.128301,Haward}. 
From a numerical perspective, the presence of stagnation points poses computational challenges, such as controlling  the exponential growth of the stress tensor (\ref{eqAdim3}) in regions close to these points. 
Some recent advances in numerical schemes for solving viscoelastic fluid flows with stagnation points, geometric singularities, and regions with a high rate of deformation have been discussed  in~\cite{FATTAL2004281,BALCI2011546,EVANS2019462,Alves}.

\begin{figure}[t!]
    \centering
   a) \includegraphics[scale=0.4]{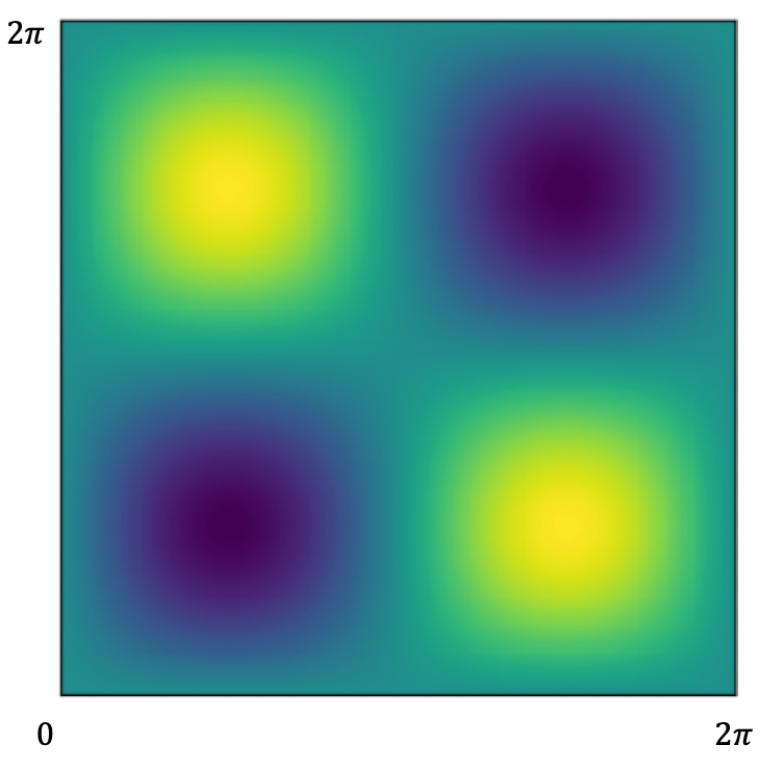} \quad 
   b)    \includegraphics[scale=0.4]{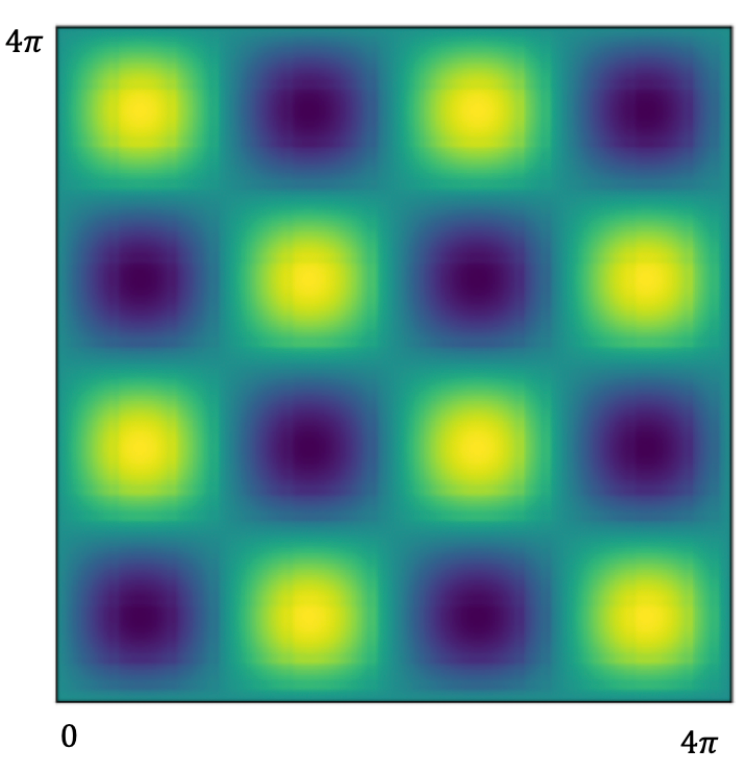}
    \caption{Vorticity plots for the four-roll mill geometry considering: a) $n=1$ and b) $n=2$.}
    \label{illus_v}
\end{figure}

In summary, we explore the four-mill benchmark, simulating the system of equations (\ref{eqAdim12})-(\ref{eqAdim3}) in a square domain $[0,(n)2\pi] \times [0,(n)2\pi]$ subject to periodic boundary conditions for all fields. 
Equations (\ref{eqAdim12})-(\ref{eqAdim3}) were solved by the open-source software Basilisk~\cite{Popinet2013-Basilisk}, which solves the mass and momentum equations using a projection scheme combined with the Bell-Collela-Glaz advection scheme in a finite-difference context. 
An important feature of this framework is the availability of the log-conformation methodology~\cite{FATTAL2004281} for solving the constitutive equations of viscoelastic models. This stabilization scheme is essential to numerically preserve the symmetric positive definite property of the conformation tensor $\textbf{C}$ during the transient fluid flow. A detailed work describing the efficiency and stability of the Basilisk code for solving transient viscoelastic fluid flows was presented in~\cite{LOPEZHERRERA2019144}.
The flow is initially driven by the constant external force $\textbf{f}=(2\sin \textbf{x} \cos \textbf{y}, -2\cos \textbf{x} \sin \textbf{y})^{T}$, with an initial conformation tensor $\textbf{C}_{t = t_0}=\boldsymbol{I}$, where $\boldsymbol{I}$ is the identity matrix. As recently discussed in~\cite{dzanic_from_sauret_2022}, the level of periodicity can be increased with the parameter $n$; in this study, we  consider $n=1$ and $n=2$, as shown in Fig. \ref{illus_v}.

\section{Reduced-order model framework}

The primary contribution of this work is to develop data-driven techniques for obtaining reduced-order models of viscoelastic flows.  In this section, we discuss how to tailor the POD and SINDy approaches to obtain reduced coordinates and dynamical systems models, respectively, for viscoelastic flows. 
 A summary of this methodology is presented in Fig. \ref{illus_frm}. In the following sections, this approach will be applied to first identify a reduced-order model for a fixed set of flow parameters, and then to identify a fully parameterized model that captures bifurcation phenomena over a range of flow parameters.

\subsection{Viscoelastic proper orthogonal decomposition}

We are interested in reduced-order models that can be computed efficiently, and therefore our first step is to perform dimensionality reduction via POD. There are few recent works that use POD for viscoelastic flow data. In~\cite{Wang}, the authors investigated turbulent drag reduction for viscoelastic fluids in a high Reynolds number channel flow. Gutierrez-Castillo and Thomases~\cite{GUTIERREZCASTILLO201948} analyzed the viscoelastic POD at zero Reynolds number for the four-roll mill geometry, where the dynamics were captured using few modes, e.g. 3-6 modes for the oscillatory regime and 14 modes for the aperiodic regime.

Following the ideas presented in~\cite{GUTIERREZCASTILLO201948}, we have defined a state-vector $\boldsymbol{q} \in \mathbb{R}^{D}$ to apply the POD decomposition, i.e.,
\begin{equation}
    \boldsymbol{q}(\textbf{x},t) = \begin{pmatrix}
        B_{xx}(\textbf{x},t)   \\
        B_{xy}(\textbf{x},t)  \\
        B_{yy}(\textbf{x},t)
    \end{pmatrix}
\end{equation}
where 
\begin{equation}
   \boldsymbol{B}  = \begin{pmatrix}
        B_{xx} & B_{xy} \\
        B_{xy} & B_{yy}
    \end{pmatrix} 
\end{equation}
is the symmetric square-root of the conformation tensor \textbf{C}. 
This choice is mathematically rooted in the fact that adopting this symmetric square-root matrix, the stored elastic energy ($E_{e}$) in a volume of fluid
can be well-defined in terms of the inner-product, i.e.
\begin{equation}
    {E}_{e} = \langle \boldsymbol{B},\boldsymbol{B} \rangle.
\end{equation}
Further details can be found in~\cite{DOERING200692}.

We perform the POD analysis using the method of snapshots~\cite{Taira} which can be summarized as:
\begin{enumerate}
    \item Using the state-vector $\boldsymbol{q}(\textbf{x},t)$ after subtracting the time-averaged field $\bar{\boldsymbol{q}}$, a data matrix $\boldsymbol{X} \in \mathbb{R}^{D\times N N_{w}}$ is constructed where $D$ is the product of the number of mesh points and the number of variables in each spatial location, while $N$ is the number of snapshots in time and $N_{w}$ is the number of parameters considered in the simulations.
    \item The following eigendecomposition is solved 
\begin{equation}
\boldsymbol{X}^{T} \boldsymbol{X} \boldsymbol{\Psi} = \boldsymbol{\Psi} \boldsymbol{\Lambda}, \\ 
\end{equation}
where $\boldsymbol{\Psi} \in \mathbb{R}^{NN_{w}\times NN_{w}} $ is a matrix whose columns are the eigenvectors and $\boldsymbol{\Lambda} \in \mathbb{R}^{NN_{w}\times NN_{w}}$ is a diagonal matrix of the eigenvalues.
\item The POD modes can be computed by
\begin{equation}
    \boldsymbol{\Phi} = \boldsymbol{X} \boldsymbol{\Psi} \boldsymbol{\Lambda}^{-1/2},
\end{equation}
with $\boldsymbol{\Phi} \in \mathbb{R}^{D\times NN_{w}} $.  The temporal coefficients $\boldsymbol{a}_{j}, \quad j=1,2,...,NN_{w}$ in time are obtained by the rows of the transposed eigenvector matrix $\boldsymbol{\Psi}^T$. 
\end{enumerate}
This procedure is also illustrated in Fig. \ref{illus_frm}.

It it possible to approximate the flow field at time $t$ from a SINDy model by taking the integrated state $\mathbf{a}(t)$ from SINDy as $\tilde{\boldsymbol{x}}(t) = \boldsymbol{\Phi}_{r} (\boldsymbol{\Lambda}_{r})^{1/2}\boldsymbol{a}(t)$; $r$ is the number of modes retained in the model.   

\subsection{System identification}
\label{Sindy}
One of the goals of decomposing $\boldsymbol{q}$ into a low-dimensional POD basis is to find a reduced-order model for the low-dimensional temporal modes $\boldsymbol{a}(t)$. 
There is a long history in fluid dynamics of building POD-Galerkin models~\cite{rowley2004model,noack2011galerkin,Benner2015siamreview,Rowley2017arfm} obtained by substituting the POD expansion of $\boldsymbol{q}$ into the governing Navier-Stokes (or in our case, the viscoelastic analogues) and integrating out the spatial degrees of freedom through orthogonal projection to obtain a differential equation for the evolution of $\boldsymbol{a}$:
\begin{align}
    \frac{d}{dt}{\boldsymbol{a}} = \boldsymbol{f}(\boldsymbol{a}).
\end{align}
These POD-Galerkin models have been used as computationally efficient reduced-order models for decades. However, this procedure is intrusive, requiring access to a working simulation where it is possible to isolate individual terms in the governing equations to compute $\boldsymbol{f}$.  Moreover, POD-Galerkin models are often unstable~\cite{carlberg2017galerkin}.

Recent data-driven techniques enable the non-intrusive learning of a model $\boldsymbol{f}$ without requiring the governing equations. 
We learn viscoelastic ROMs with the sparse identification of nonlinear dynamics (SINDy)~\cite{Brunton} method, which uses sparse regression to learn a model $\boldsymbol{f}$ with the fewest terms possible from a library of candidate functions that might describe the dynamics. 

The SINDy regression problem may be written as a sparse optimization problem according to the following algorithm
\begin{enumerate}
    \item A library  of candidate functions $\boldsymbol{\Theta}(\boldsymbol{a},w)$ is constructed assuming that $\boldsymbol{\dot{a}} \approx  \boldsymbol{\Theta}(\boldsymbol{a},w)\boldsymbol{\Xi}$ is a good approximation for $\boldsymbol{f}$ for a sparse matrix of coefficients $\boldsymbol{\Xi}$. 
    This provides flexibility to learn the dynamics when the functional form of $\boldsymbol{f}$ is unknown. For our library, we consider linear and cubic functions of the first two modes $\boldsymbol{a}_{1}$ and $\boldsymbol{a}_{2}$ including the influence of a nondimensional parameter $w=\frac{1}{Wi}$, the inverse of the Weissenberg number which describe the elastic effects in viscoelastic instabilities.
    \item A sparse optimization problem is now defined to solve for the sparse matrix of coefficients $\boldsymbol{\Xi}$ that selects the terms from the library that are active in the dynamics:
    \begin{equation}
    \text{argmin}_{\boldsymbol{\Xi}} || \boldsymbol{\dot{a}}- {\boldsymbol{\Theta}(\boldsymbol{a},w)}\boldsymbol{\Xi} ||_{2}^{2} + \gamma R(\boldsymbol{\Xi}), 
        \label{opt2}
    \end{equation}
    where $R(\boldsymbol{\Xi})$ is a regularization term (e.g. $l_{1}$ norm for the current work); $\gamma$ is a hyperparameter for the regularization. 
    \item The optimization problem (\ref{opt2}) is then solved by the Sparse relaxed regularized regression (SR3) algorithm~\cite{zheng2018unified,champion2020unified} using the open-source PySINDy code~\cite{desilva2020,Kaptanoglu2022} in order to obtain the dynamics of $\boldsymbol{a}(t)$.
    
\end{enumerate}
For tests using a fix nondimensional parameter, i.e. $N_{w}=1$, the library is simplified as
\begin{equation}
{\boldsymbol{\Theta}(\boldsymbol{a})} = 
\left[
  \begin{array}{cccc}
    \vertbar & \vertbar &    \vertbar \\
    \textbf{1}    & \boldsymbol{a}    &(\boldsymbol{a} \otimes  \boldsymbol{a}  \otimes\boldsymbol{a})    \\
    \vertbar & \vertbar   &\vertbar 
  \end{array}
\right]
\label{theta_a}
\end{equation}
while to include the influence of the control parameter $w$ we have
\begin{equation}
{\boldsymbol{\Theta}(\boldsymbol{a},w)} = 
\left[
  \begin{array}{cccc}
    \vertbar & \vertbar &    \vertbar \\
    \textbf{1}    & \boldsymbol{a}    &(\boldsymbol{a} \otimes  \boldsymbol{a}  \otimes\boldsymbol{a})    \\
    \vertbar & \vertbar   &\vertbar 
  \end{array}
\right]
\otimes
\left[
  \begin{array}{cccc}
    \vertbar & \vertbar & \vertbar  &  \vertbar \\
    1    & w  & w^2  &w^3    \\
    \vertbar & \vertbar & \vertbar  &\vertbar 
  \end{array}
\right].
\label{theta_a2}
\end{equation}

Typically a hyperparameter scan is required to find the model that best balances the tradeoff between training accuracy and generalizability. The optimum value for the regularization parameter considered in this study was determined by hyper-parameter tuning to be $\gamma=10^{-4}$.
 
\subsection{Model stability}
\label{Stab}
ROMs do not generally come with guarantees that their predictions of the dynamics will stay bounded, even if the original system has global boundedness~\cite{carlberg2017galerkin}.  One advantage of the identified SINDy models here is that a nonlinear stability analysis illustrates the conditions under which the resulting system is stable for any initial condition $\boldsymbol{a}(0)$, i.e. the model is long-time bounded for any initial condition~\cite{schlegel2015long,kaptanoglu2021promoting}. Using $K = \frac{1}{2}\boldsymbol{a} \cdot \boldsymbol{a} \geq 0$ as a Lyapunov function, we can construct the relation
\begin{align}
    \label{eq:energy_stability}
    \dot{K} = \boldsymbol{a} \cdot \dot{\boldsymbol{a}}. 
\end{align}
For boundedness, a sufficient condition is that $\dot{K} < 0$ for $\|\boldsymbol{a}\|_2$ large enough.   For large $\|\boldsymbol{a}\|_2$, and assuming there are nonzero cubic terms in the model, we need only consider the quartic terms in the modes energy since they are dominant in this regime. 

\section{Results for a fixed Weissenberg number} \label{SecR1}

We first explore the use of SINDy to develop reduced-order models for viscoelastic fluids at a fixed Weissenburg number.  The majority of applications of SINDy in fluids have developed models for a fixed set of parameters, with some notable exceptions~\cite{deng2020low,deng2021galerkin}.  Developing a model for a fixed parameter (e.g., so that $N_{w}=1$) is typically simpler, making it a reasonable starting point for our analysis.  

Simulations were performed using a uniform mesh size of $128 \times 128$ for the two levels of periodicity $n=1$ and  $n=2$ and the time-step was $\delta t =0.0001$ for all cases. 
To capture the essential transition from a steady state flow to an oscillatory regime, we run the simulation for a long nondimensional time, i.e., $t_{end}=750$. We fix the Reynolds number at $Re=1$ to reduce the inertial effects during the transient flow, allowing us to focus on the onset of viscoelastic instabilities. 
The mesh and Reynolds number used here are similar those in~\cite{dzanic_from_sauret_2022}. 
In addition, consistent with previous numerical studies~\cite{GUTIERREZCASTILLO201948,PhysRevE.106.L013101}, we use $\beta Wi<0.5$ for all simulations. As commented in~\cite{LIU201238}, the viscosity ratio $\beta$ is another important viscoelastic parameter that can influence the flow transitions. 
The recent work of~\cite{PhysRevLett.129.017801} demonstrated that the flow transitions in viscoelastic systems, with purely elastic flow structures, can be numerically captured with $\beta$ in the range $0.6\leq \beta \leq 0.95$. In addition, there is an important connection between the elastic and viscous forces~\cite{THOMPSON2021104550}, which can be captured changing the Weissenberg number as well as the viscosity ratio. Since high values of $\beta$ contribute to the stability of the method, we have fixed $\beta=0.9$. 

\begin{figure}[t!]
\vspace{-.2in}
    \centering
   a) \includegraphics[scale=0.5]{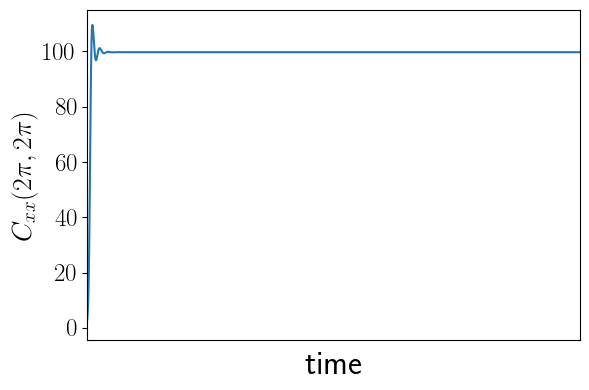} \quad 
   b)  \includegraphics[scale=0.5]{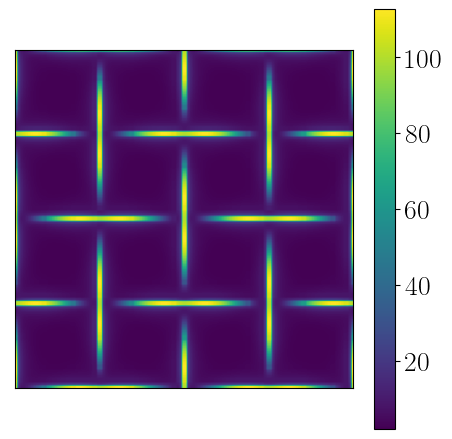}  \\
c) \includegraphics[scale=0.5]{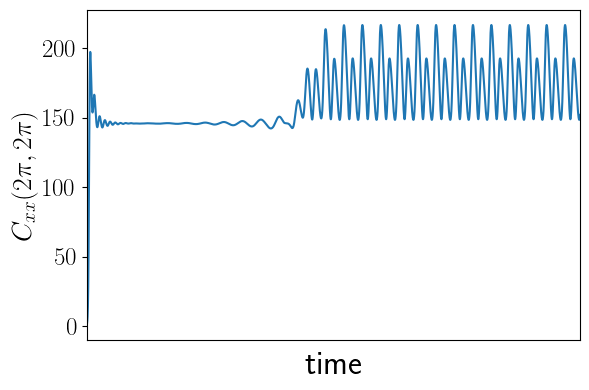} \quad 
d)    \includegraphics[scale=0.5]{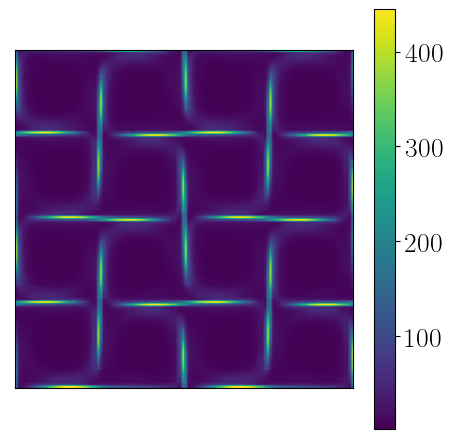} 
      \caption{Temporal variation of the $C_{xx}$-component of the conformation tensor at the central stagnation point of the four-roll mill geometry with $n=2$: a) $Wi=2$ and c) $Wi=4$. Results for spatial distributions of the trace of the conformation tensor $tr(\textbf{C})$ at time $400$ are presented in b) for $Wi=2$ and d) $Wi=4$.}
    \label{illus1}
\end{figure}

To illustrate the elastic effects on the four-roll mill geometry, two simulations with $n=2$ are shown in Fig. \ref{illus1}. 
This figures shows the transient behaviour of the first component of the conformation tensor $\textbf{C}$, denoted here as $C_{xx}$, for a sensor located at the central stagnation point of the domain. 
For a low Weissenberg number ($Wi=2$), the flow becomes steady after the first transient peak, while for $Wi=4$ there is a transition of the flow for into an oscillatory regime. These results are qualitatively in agreement with those presented in~\cite{GUTIERREZCASTILLO201948,dzanic_from_sauret_2022}, e.g., the choice of the Weissenberg number is critical to characterize the long-time dynamics and bifurcations from steady symmetries to oscillatory regimes in this geometry. We have limited the Weissenberg number to $Wi<5$ in our study to avoid elastic turbulence, which is beyond the scope of this work. 
Figure~\ref{illus1} also shows the snapshots of the trace of the conformation tensor $\textbf{C}$, denoted here as $tr(\textbf{C})$, at the final simulation time. The trace of the conformation tensor is frequently used in the viscoelastic literature~\cite{Alves} to quantify the elastic energy of the system. For $Wi=2$ the flow is steady with symmetric patterns for $\textbf{C}$ while for $Wi=4$ the flow becomes asymmetric, resulting in an oscillatory regime. 
When computing a POD basis, we use data after the first peak~\cite{GUTIERREZCASTILLO201948} so that our reduced-order models capture steady-state or long-time dynamics. 

\begin{figure}[t!]
    \centering
   \includegraphics[scale=0.56]{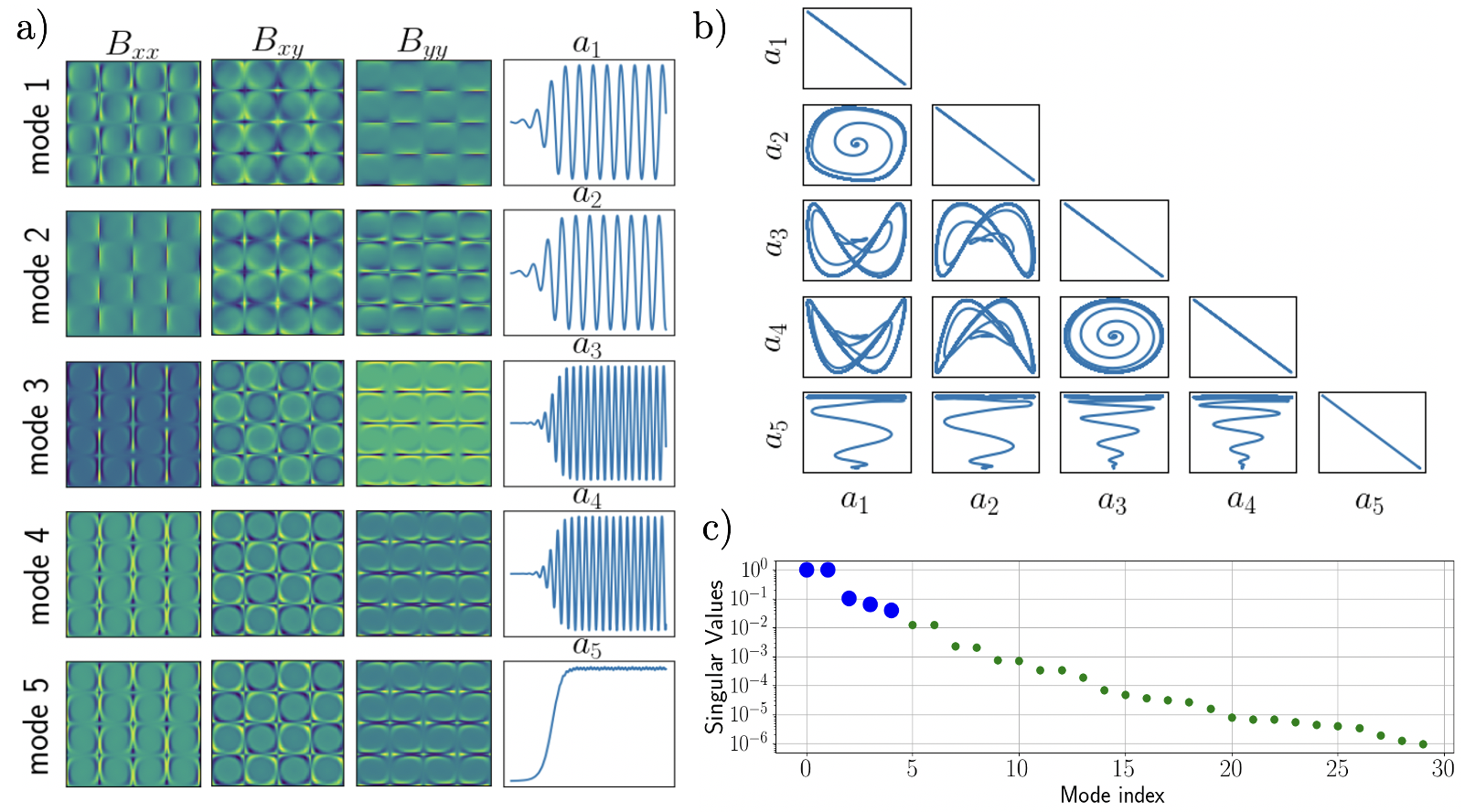} 
    \caption{POD analysis for four-roll mill geometry considering two levels of periodicity ($n=2$): a) First five spatial modes for $B_{xx}, B_{xy}$ and $B_{yy}$ and the first five temporal coefficients $a_i, i=1,...,5$, b) mode pair trajectories and c) normalized singular values.}
    \label{illus_frm3}
\end{figure}
We first simulate a Weissenberg number of $Wi=3.5$ for $n=1$ and $n=2$. The POD analysis is shown for $n=2$ in Fig.~\ref{illus_frm3}, where we plot the first five spatial and temporal modes of $\boldsymbol{B}$. The phase portraits are also shown, and the results for $n=1$ are omitted since they are similar.

Using the framework in Section \ref{Sindy}, the learned SINDy model is
\begin{align}
\begin{pmatrix}
        \dot{a}_{1}   \\
        \dot{a}_{2} 
    \end{pmatrix} 
    = \begin{pmatrix}
        \epsilon_1 & \epsilon_2 \\
        -\epsilon_2 & \epsilon_1
    \end{pmatrix}
    \begin{pmatrix}
        {a}_{1}   \\
        {a}_{2} 
    \end{pmatrix}  
+
\begin{pmatrix}
        \delta_1{a}_{1}^{2}+\delta_2{a}_{2}^{2}+\delta_3{a}_{1}{a}_{2}  & \delta_4{a}_{2}^{2} \\
        -\delta_4{a}_{1}^{2} & \delta_2{a}_{1}^{2}+\delta_1{a}_{2}^{2}-\delta_3{a}_{1}{a}_{2} 
    \end{pmatrix}
    \begin{pmatrix}
        {a}_{1}   \\
        {a}_{2} 
    \end{pmatrix}, 
    \label{si}
\end{align}
where the coefficients $\epsilon_1$, $\epsilon_2$, $\delta_1$,  $\delta_2$,  $\delta_3$ and $\delta_4$ vary according the fluid parameters and periodicity $n$. 
\begin{table}[!t]
    \centering
    \caption{Coefficients for the sparse model (\ref{si}) for $n=1$ and $n=2$.}
    \begin{tabular}{c c c c c c c }
    \hline 
    Temporal interval &   $\epsilon_1 $ & $\epsilon_2 $ & $\delta_1 $ &   $\delta_2 $ &   $\delta_3 $&   $\delta_4 $ \\
    \hline 
   $t \in [250,750]$ ($n=1$)     &  $0.006 $     &   $ -0.023$         &   $ -0.028$        &   $ -0.008$    &   $ 0.038$         &   $ -0.016$  \\
   $t \in [200,750]$ ($n=2$)      &  $0.007 $     &   $ 0.027$  &   $ -0.045$        &   $ 0.023$    &   $ -0.026$         &   $ 0.016$  \\
     \hline
 
    \end{tabular}
    \label{ttw}
\end{table}

\subsection{Stability of the system}

To assess the stability of system~\eqref{si}, we investigate the conditions for $\dot{K} < 0$:
\begin{align}
    \dot{K} \propto \delta_1(a_1^4 + a_2^4) + 2\delta_2a_1^2a_2^2 + (\delta_3-\delta_4)(a_1^3a_2 - a_2^3a_1).
\end{align}
It is now clear that $\delta_1 \leq 0$ is a sufficient condition for $\dot{K} < 0$ for large $\|\boldsymbol{a}\|_2$ by considering $a_1(t) = 0$ and $a_2(t) \neq 0$, or vice-versa. 
Likewise, we conclude that a sufficient condition is $\delta_1 + \delta_2 < 0$ from the line $a_1(t) = a_2(t)$. With these conditions,  we see that the term proportional to $(\delta_3-\delta_4)$ remains subdominant for large enough $\|\boldsymbol{a}\|_2 \gg 1$ and therefore there is no boundedness-related restriction on these coefficients.  

To summarize, for large enough $a_1 \gg 1$, $a_2 \gg 1$, these conditions on $\delta_1$ and $\delta_2$ guarantee that the temporal derivative of the energy is negative, meaning all trajectories fall into a monotonically trapping region, similar to that found for some quadratic fluid models~\cite{schlegel2015long,kaptanoglu2021promoting}. These conclusions are unchanged when the Weissenberg number is later included in the model as a control parameter, since the new terms only provide low-order contributions to the energy that are irrelevant far from the origin. Reduced-order models built satisfying these constraints can be trusted to produce bounded predictions of the dynamics even with new and arbitrary initial conditions and arbitrary time intervals.

\subsection{Predictions of the POD coefficients}

Table~\ref{ttw} shows the coefficients of the reduced-order model (\ref{si}) over the temporal interval $t \in [250,750]$ for $n=1$ and $t \in [200,750]$ for $n=2$ resulting in $N=5000$ and $N=5500$ snapshots, respectively.  
The coefficients of the linear terms are close for both values of $n$. However, the level of periodicity affects the variation of the cubic terms. 
These results suggest that the level of periodicity could be included as an additional parameter in the reduced-order model.

The reduced-order model (\ref{si}) was verified to accurately forecast the first two temporal coefficients $a_1$ and $a_2$ for both values of periodicity, $n=1$ and $n=2$. 
We further improved the model by learning a sparse algebraic relationship for the higher harmonic modes: 
\begin{equation}
    a_{i} = g(a_1,a_2), \quad i=3,4.
    \label{rel1}
\end{equation}
Using a quadratic library constructed from $a_1$ and $a_2$, we obtain the following function $g$
\begin{equation}
   a_{i}= \gamma_1 a_1a_2 + \gamma_2(a_{1}^{2} - a_{2}^{2}), \quad i=3,4,
    \label{rel2}
\end{equation}
where $\gamma_1$ and $\gamma_2$ are given in Table \ref{ttw22} for $n=1$ and $n=2$.

\begin{table}[!t]
    \centering
    \caption{Coefficients for $a_3$ and $a_4$ for the resulting nonlinear equation (\ref{rel2}).}
    \begin{tabular}{c c c}
    \hline 
    $a_i$ (level of periodicity $n$) &   $\gamma_1 $ & $\gamma_2 $ \\
    \hline 
    $a_3$ ($n=1$)       &  $-2.73 $     &   $-1.45$ \\
    $a_4$ ($n=1$)       &  $2.38 $     &   $ -1.53$ \\
    $a_3$ ($n=2$)       &  $-3.18 $     &   $1.39$ \\
    $a_4$ ($n=2$)       &  $2.07 $     &   $ 1.58$ \\
     \hline
    \end{tabular}
    \label{ttw22}
\end{table}

\begin{figure}[t!]
\vspace{-.1in}
    \centering
a)\includegraphics[width=.9\textwidth]{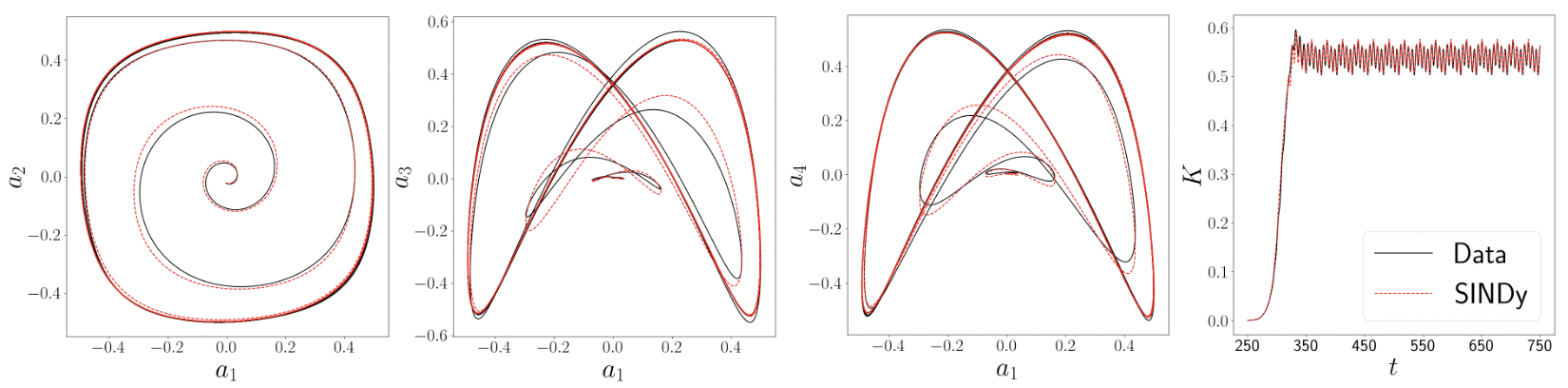} \\
b)\includegraphics[width=.9\textwidth]{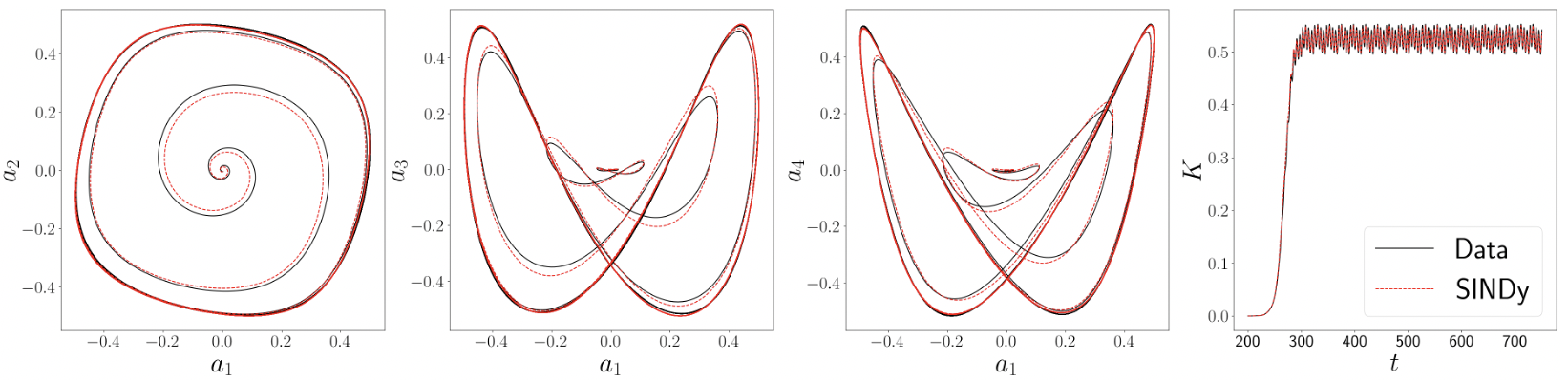}
\vspace{-.2in}
   \caption{Comparison of the mode pair trajectories in time considering the pairs $(a_1,a_2)$, $(a_1,a_3)$ and $(a_1, a_4)$ and the quantity $K = \frac{1}{2}\boldsymbol{a} \cdot \boldsymbol{a}$ value using the full numerical solution (Data) and the reduced-order model (SINDy) for four-roll mill geometry: a) $n=1$ and b) $n=2$.}
    \label{4Rcomp3}
\end{figure}

Figure~\ref{4Rcomp3} shows the trajectories of \eqref{si} in phase space.  
The trajectories of the identified systems in phase planes $(a_1,a_2)$, $(a_1,a_3)$ and $(a_1,a_4)$ are in excellent agreement with those of the full-order model simulations. 
As a further verification, the temporal evolution of the quantity $K = \frac{1}{2}\boldsymbol{a} \cdot \boldsymbol{a}$, based on the first four modes, is also shown in Fig.~\ref{4Rcomp3}, thus confirming that stability is preserved in the model.

\begin{figure}[t!]
    \centering
\includegraphics[width=.875\textwidth]{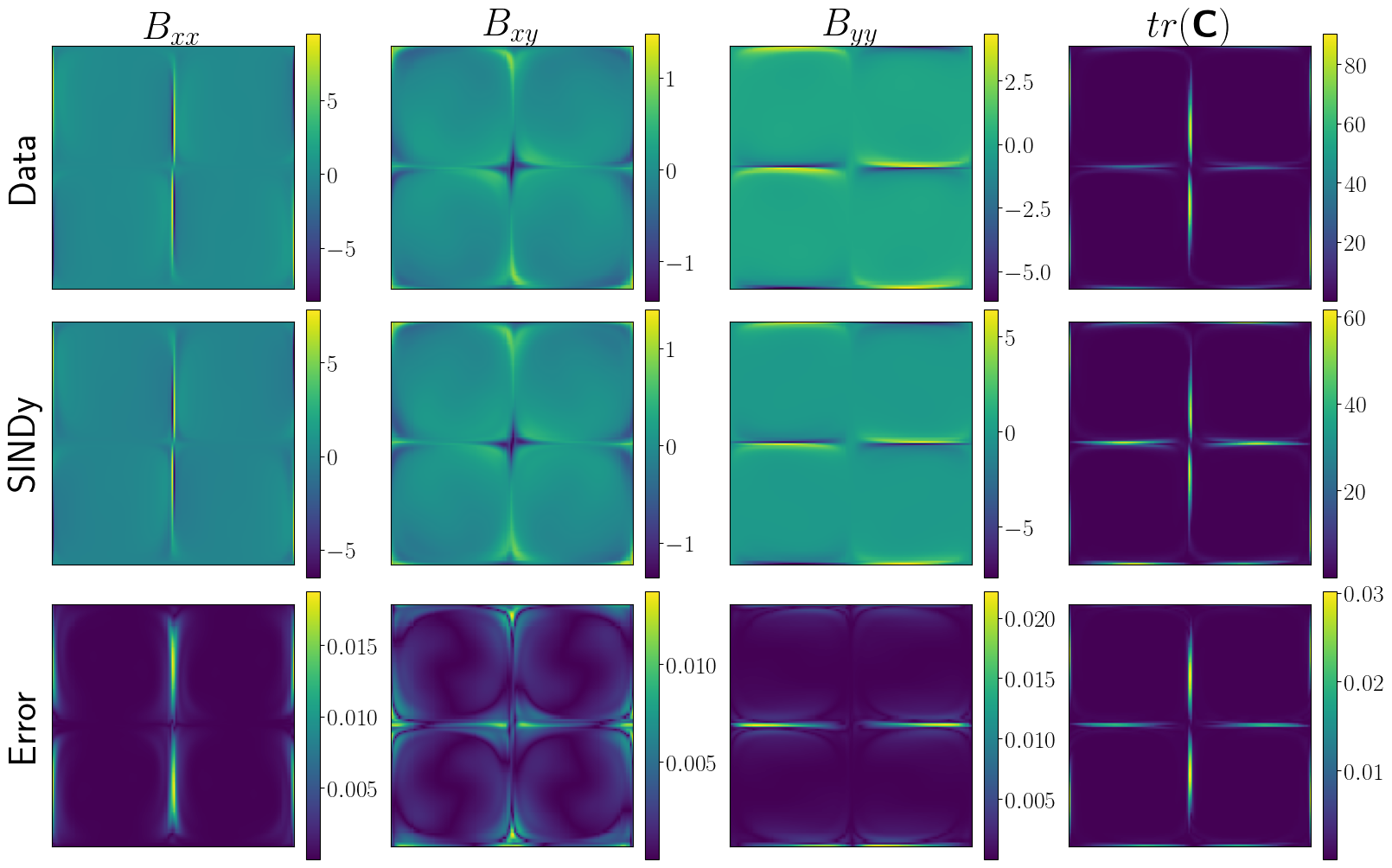}
\vspace{-0.1in}
\caption{Comparison between the  full numerical solution (Data), the reduced-order model (SINDy), and the error for $n=1$ at simulation time $t=750$.}  
    \label{4Rrecons3}
\end{figure}

\subsection{Flow field reconstruction}

\begin{figure}[t!]
    \centering
    \vspace{-.15in}
\includegraphics[width=.875\textwidth]{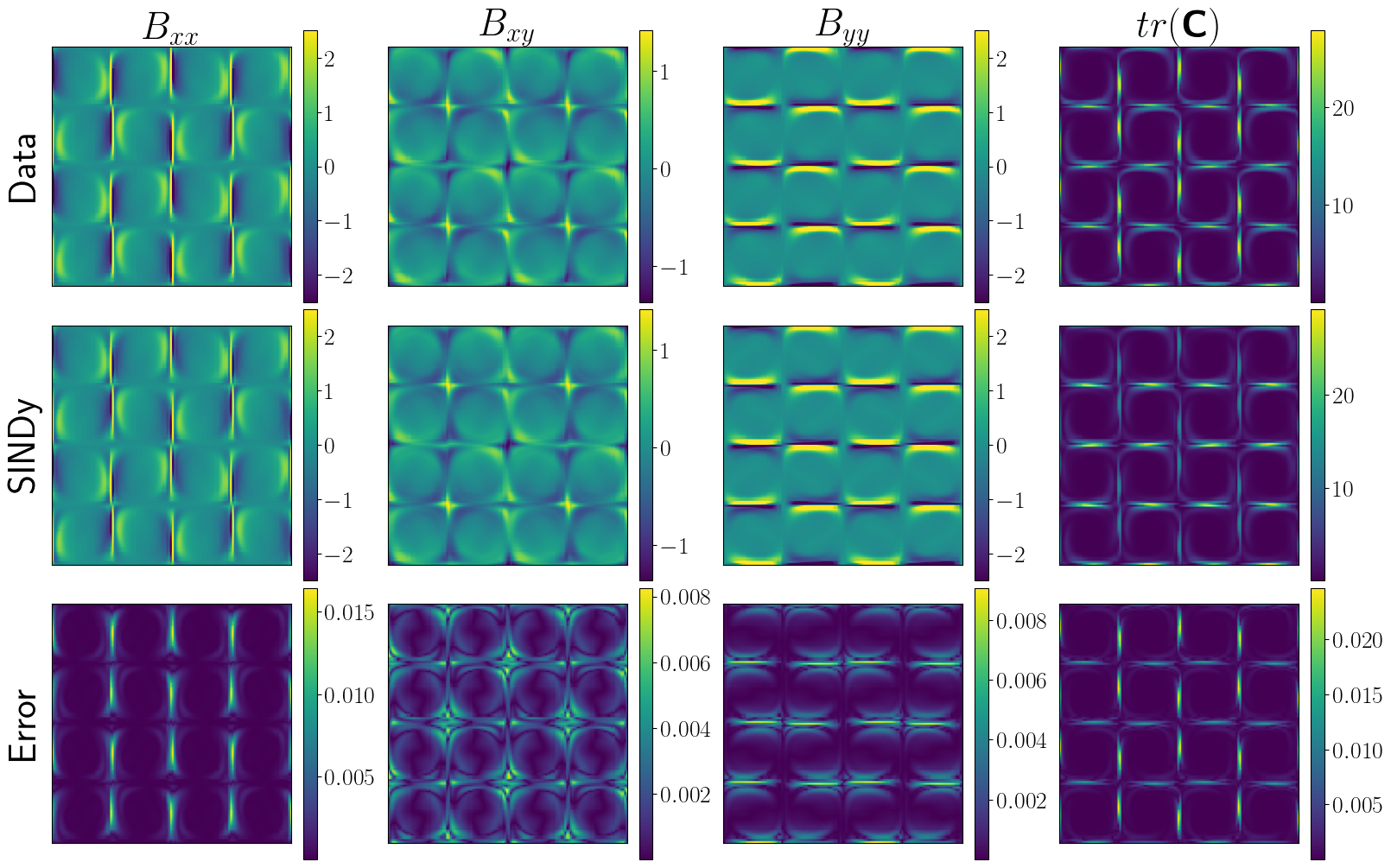}
\vspace{-0.1in}
\caption{Comparison between the  full numerical solution (Data), the reduced-order model (SINDy), and the error for $n=2$ at simulation time $t=750$.}  
    \label{16Rrecons3}
\end{figure}

Finally, we reconstruct the high-dimensional flow field from the low-dimensional model using POD modes. Figures \ref{4Rrecons3} and \ref{16Rrecons3} show the flow field of the full order model (Data), the reconstructed flow fields using the reduced-order model (SINDy) and the normalized error in the $2-$norm, for $n=1$ and $n=2$ respectively. 
In particular, these figures reconstruct the components of the symmetric square-root of the conformation tensor $\boldsymbol{B}$ and the trace of the conformation tensor $\textbf{C}$, which can be computed as
\begin{equation}
   tr(\textbf{C}) =  B_{xx}^{2} + B_{yy}^{2} + 2 B_{xy}^{2}.
\end{equation}
According to the results in Figs. \ref{4Rrecons3} and \ref{16Rrecons3} we observe the effectiveness of the proposed framework to reconstruct the main signatures of the flow for all components of $\boldsymbol{B}$ as well as for $tr(\textbf{C})$. From a quantitative standpoint, the errors remain below $\%3$ in all cases. As expected, the highest errors are  concentrated around the vicinity of the stagnation points.

\clearpage
\section{Results for system parameterized by Weissenberg number} 
 \label{SecR2}

\begin{figure}[t!]
    \centering
   a) \includegraphics[scale=0.5]{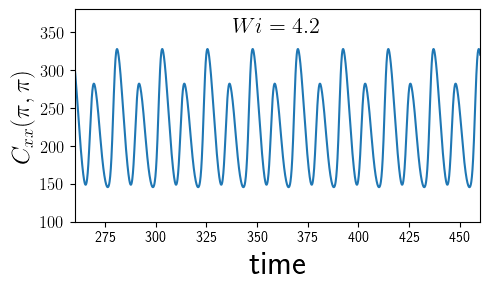} \quad 
   b)    \includegraphics[scale=0.5]{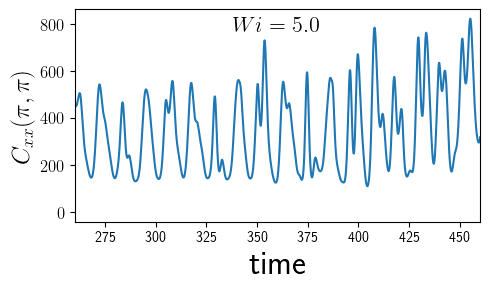}
    \caption{Time series of the $C_{xx}$-component of the conformation tensor at the central stagnation point of the four-roll mill geometry with $n=1$: a) Oscillatory regime and b) Aperiodic regime.}
    \label{oscill_chaotic}
\end{figure}

As a challenging test case, we develop a SINDy model that is parameterized by the Weissenberg number $Wi$ and show that this model is valid over a wide range of values.  
In particular, we apply SINDy with control~\cite{Kaiser2018prsa,Fasel} using the library \textbf{$\Theta(\boldsymbol{a},w)$} with the control input defined as $w={1}/{Wi}$. 
For the roll-mill simulations with $n=1$, {we train on the values $Wi=\{4,4.35,4.5\}$ and test on the values $Wi=4.2$ and $Wi=5.0$. The $Wi$-values chosen for training are specifically selected to maintain the flow in an oscillatory regime. Consequently, the dynamics for the interpolation value $Wi=4.2$ also exhibit the same oscillatory behavior, as evident in Fig. \ref{oscill_chaotic}a). However, for $Wi=5.0$, the flow undergoes a dynamic transition to chaotic behavior, driven by increasing elastic effects, as illustrated in Fig. \ref{oscill_chaotic}b). This test value is chosen to demonstrate the extrapolation capabilities of our model, particularly in capturing the transition from an oscillatory regime to a chaotic one. The increase in Weissenberg number also leads to a significant increase in the first component of the conformation tensor at the stagnation point, shown in Fig. \ref{oscill_chaotic}, aligning with findings from prior studies (see \cite{dzanic_from_sauret_2022}).} For the sake of computational efficiency, we use a uniform mesh size of $64 \times 64$ for this dataset. The temporal window is $t \in [260, 460]$ resulting in $N=2000$ snapshots.  We also test this parametric SINDy approach for $n=2$ training on $Wi=\{3.6,4\}$ and testing on $Wi=3.8$, with $N=800$ snapshots.

SINDy results in a dynamical system for the first two temporal coefficients that includes the effect of the Weissenberg number. The identified system is
\begin{equation}
\begin{split}
\begin{pmatrix}
        \dot{a}_{1}   \\
        \dot{a}_{2} 
    \end{pmatrix} 
&    = \begin{pmatrix}
        \epsilon_1 & \epsilon_2 \\
        -\epsilon_2 & \epsilon_1
    \end{pmatrix}
    \begin{pmatrix}
        {a}_{1}   \\
        {a}_{2} 
    \end{pmatrix} 
\\
&+
\begin{pmatrix}
        \delta_1{a}_{1}^{2}+\delta_2{a}_{2}^{2}+\delta_3{a}_{1}{a}_{2}  & \delta_4{a}_{2}^{2} & \delta_5{a}_{1}w +  \delta_6{a}_{2}w + \delta_7w^{2}\\
        -\delta_4{a}_{1}^{2} & \delta_2{a}_{1}^{2}+\delta_1{a}_{2}^{2}-\delta_3{a}_{1}{a}_{2} & -\delta_6{a}_{1}w +  \delta_5{a}_{2}w - \delta_7w^{2}
    \end{pmatrix}
    \begin{pmatrix}
        {a}_{1}   \\
        {a}_{2}   \\
        w   \\
    \end{pmatrix}, 
    \label{si2}
\end{split}
\end{equation}
where $w={1}/{Wi}$. The coefficients of this ROM (\ref{si2}) are in Table \ref{ttw3}, and the coefficients of the algebraic expression for the higher harmonics in Eq. (\ref{rel2}) are in Table \ref{ttw23-1}. 
The qualitative conclusions are the same for $n=1$ and $n=2$, so we only describe the results for $n=1$. 

\begin{table}[!t]
    \centering
    \caption{Coefficients of model (\ref{si2}) with training set $Wi=\{4,4.35,4.5\}$ for $n=1$ and $Wi=\{3.6,4\}$ for $n=2$.}
    \vspace{-.1in}
    \begin{tabular}{c c c c c c c c c c}
    \hline 
    level of periodicity &   $\epsilon_1 $ & $\epsilon_2 $ & $\delta_1 $ &   $\delta_2 $ &   $\delta_3 $&   $\delta_4 $ &   $\delta_5 $ &   $\delta_6 $&   $\delta_7 $\\
    \hline 
    $n=1$     &  $0.05 $     &   $-0.08$  &   $ -0.16$&   $ -0.09$    &   $ 0.004$  &   $ 0.175$ &   $ -0.57$    &   $ 0.58$         &   $ 0.01$ \\
    $n=2$     &  $0.03 $     &   $0.08$  &   $-0.04$&   $ -0.13$    &   $ -0.04$  &   $ -0.10$ &   $ -0.36$    &   $ -0.54$         &   $ 0.04$ \\
     \hline
 
    \end{tabular}
    \label{ttw3}
\end{table}

\begin{table}[!t]
    \centering
    \caption{Coefficients for $a_3$ and $a_4$ for the nonlinear equation (\ref{rel2}) using the training set $N_{w}^{train}=3$, $Wi=\{4,4.35,4.5\}$ for $n=1$ and $N_{w}^{train}=2$, $Wi=\{3.6,4\}$ for $n=2$.}
        \vspace{-.1in}
    \begin{tabular}{c c c}
    \hline 
    temporal mode (level of periodicity) &   $\gamma_1 $ & $\gamma_2 $ \\
    \hline 
    $a_3$ ($n=1$)       &  $4.80 $     &   $-0.93$ \\
    $a_4$ ($n=1$)       &  $2.87 $     &   $ 2.1$ \\
    $a_3$ ($n=2$)       &  $3.06 $     &   $ 1.93 $ \\
    $a_4$ ($n=2$)       &  $-3.97 $     &   $ 1.0 $ \\
     \hline
    \end{tabular}
    \label{ttw23-1}
\end{table}

\begin{figure}[t!]
	\centering
	\includegraphics[scale=0.4]{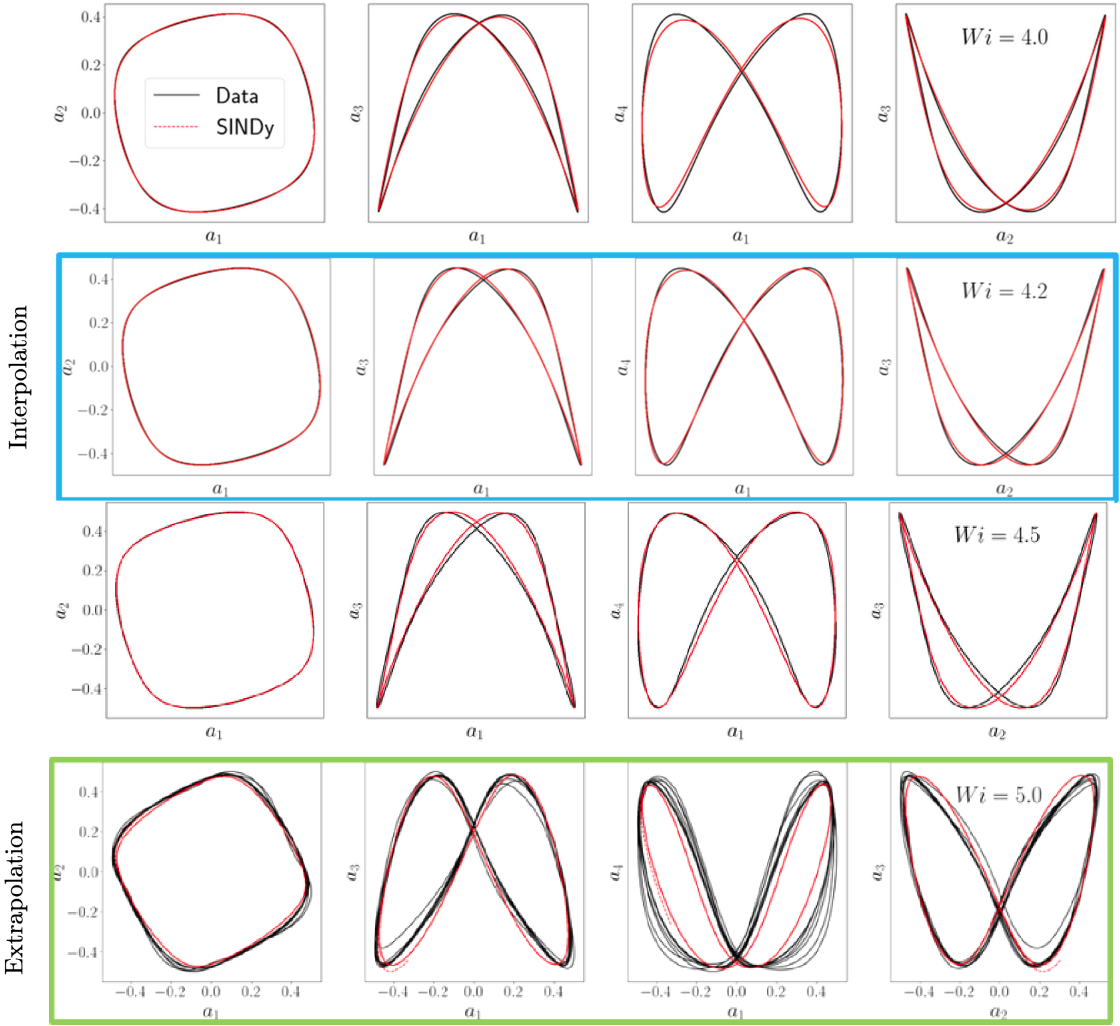}
\caption{{Comparison of the phase portraits for the pairs $(a_1,a_2)$, $(a_1,a_3)$, $(a_1, a_4)$ and $(a_2, a_3)$ using the full numerical solution (Data) and the reduced-order model (SINDy) for $n=1$ with training data $Wi=\{4,4.35,4.5\}$, with an interpolated value of $Wi=4.2$ (highlighted in blue) and an extrapolated value of $Wi=5$ (highlighted in green) for testing data.}}
\label{4Rcomp3-1}
\vspace{-.2in}
\end{figure}

The phase portraits from numerical simulations and from the SINDy model are shown in Fig. \ref{4Rcomp3-1} for  training and testing cases, where the SINDy model accurately captures the unsteady limit cycle behavior. The SINDy reduced-order model also exhibits a reasonable qualitative prediction for the extrapolated value $Wi=5$. 

\begin{figure}[t!]
    \centering
\includegraphics[width=0.85\textwidth]{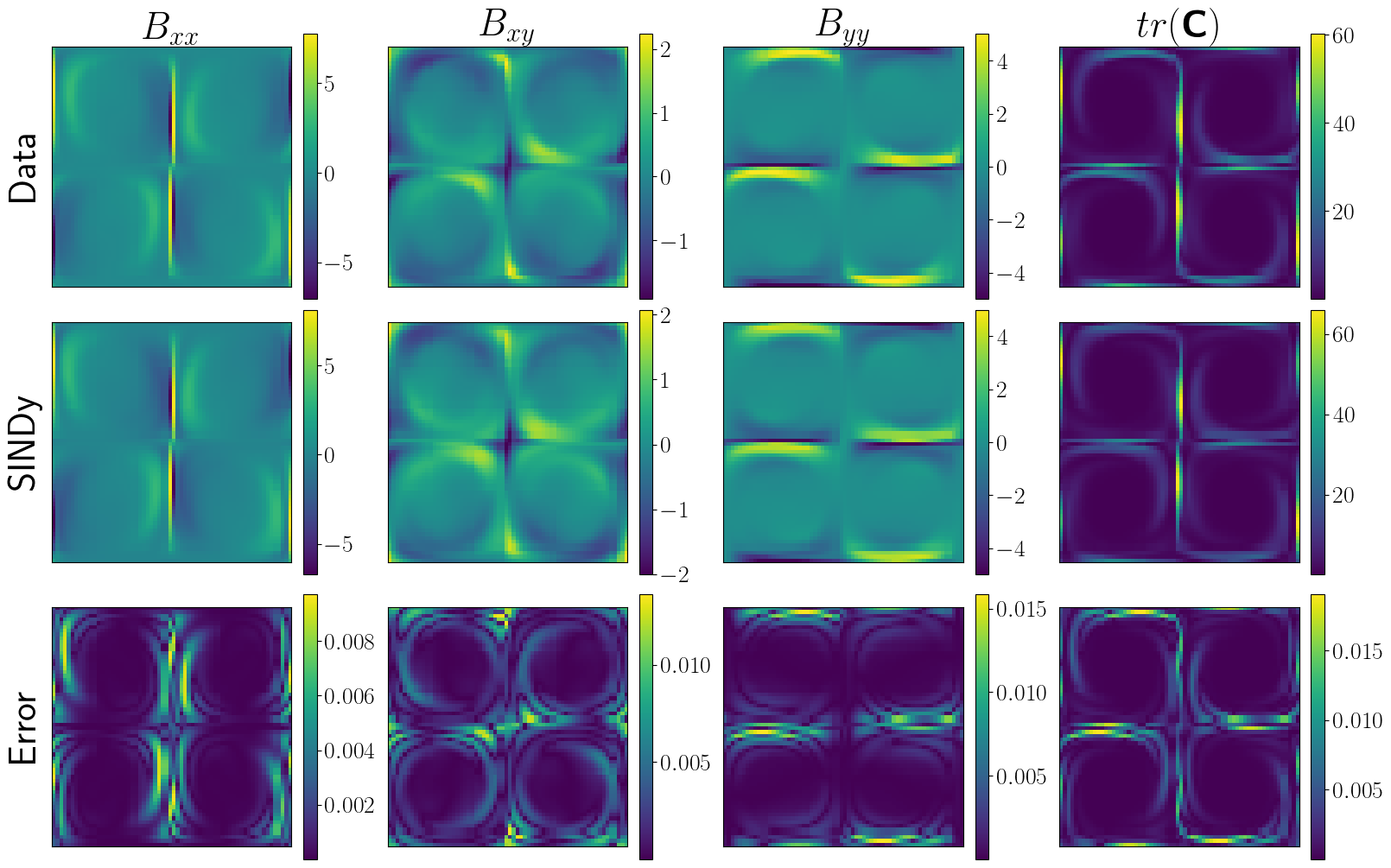} 
\caption{Comparison between the  full numerical solution (Data), the reduced-order model (SINDy), and the error for $n=1$ and $Wi=4.2$ at simulation time $t=360$.}  
    \label{4Rrecons4-1}
\end{figure}

Figure \ref{4Rrecons4-1} shows the full flow field reconstruction from the SINDy model for the testing case of $Wi=4.2$.  The SINDy model coefficients are multiplied by POD modes to obtain the full fields for the components of $\boldsymbol{B}$ and the trace of the conformation tensor. 
The reconstruction of the ROM is in excellent agreement with the fields from high-fidelity simulation, even though this was for a testing case not considered in our training data. Notably, for $Wi=4.2$, the largest reconstruction error is approximately $1.75\%$. For the extrapolated value $Wi=5$, we observe an increase in the largest error to approximately $10\%$ (not shown here for the sake of space). This is attributed to an underestimation of the transition dynamics, since the ROM uses only the first two temporal modes. More results considering high Weissenberg numbers are presented in the Appendix.

\section{Discussion}
This work explores the sparse reduced-order modeling of viscoelastic fluid flows with the SINDy algorithm. 
{SINDy has found extensive applications in fluid dynamics, yet its application to reduced-order modeling of viscoelastic flows is still in its early stages. This study is a first investigation in this domain, and serves to showcase the ability of sparse modeling to capture complex, canonical viscoelastic fluid dynamics. } Importantly, the models developed are stable by construction, interpretable, and highly structured, possessing symmetries in the model coefficients. Further, we show that these models can be used to extrapolate beyond the training data.

In addition, we construct models that are explicitly parameterized by the Weissenberg number, which controls the dominant bifurcations in the flow. {Most SINDy models in the literature are not explicitly parameterized by a dominant non-dimensional number, which is a more challenging modeling problem. Our findings indicate that these parameterized models are accurate both for interpolation and limited extrapolation tasks. Moreover, based on our initial tests at higher Weissenberg numbers (see Appendix), it appears necessary to expand the ROM dimension to accurately predict transitions and behaviors associated with elastic turbulence ~\cite{Groisman,Victor}, which is a subject of ongoing work.} 

Here we consider a canonical viscoelastic configuration using the Oldroyd-B model in the four-roll mill benchmark geometry. 
 Our modeling strategy begins with the viscoelastic POD to reduce the dimensionality and extract a few dominant viscoelastic coherent structures in the flow.  
 We then model the dynamics of the first leading POD modes with SINDy and develop algebraic for the higher harmonic modes in terms of these driving modes.  
 The comparisons conducted between the simulation data and the results obtained from our surrogate models serve as compelling evidence, confirming the stability and accuracy of our ROMs. 
 Therefore, these models provide precise predictions while requiring only minimal computational resources. 
Further, we show that accurate full flow field reconstructions are possible by recombining the POD modes using the mode coefficients predicted by the SINDy model.  
 
We also discuss the interpretability of the model, as this is a crucial challenge for data-driven system identification. Motivated by the POD-Galerkin models, our formulation resulted in a compact and interpretable dynamical system model directly from the simulation data of an oscillatory viscoleastic fluid flow. 
This work also investigates the performance of our model at higher Weissenberg numbers that were not included in the training dataset.  The SINDy model shows reasonable qualitative agreement on this extrapolation task, with a maximum reconstruction error below $10\%$; however, future work will investigate how to improve the model in the chaotic regime. 

There are a number of future directions based on this work. We considered the Oldroyd-B model, although this model fails for extensional flows at high Weissenberg number~\cite{PhysRevFluids.7.080701}. A natural extension is the development of ROMs for nonlinear viscoelastic models, such as the Phan-Thien and Tanner (PTT) and Finitely Extensible Nonlinear Elastic (FENE-P) fluids, in non-viscometric flows, for instance cross-slot geometry~\cite{Alves}. 
Further, it will be interesting to include the parametric effect of viscosity ratio and extensibility factor in addition to the Weissenberg number. 
Because our strategy exhibits exponential scaling with the number of parameters, addressing multiparameter viscoelastic models (such as PTT and FENE-P) will require more sophisticated SINDy optimization algorithms. This becomes particularly crucial for enhancing computational efficiency in future models.
Ongoing work by the authors will also investigate the asymptotic behavior of the model. Finally, data-driven investigations of more complex non-Newtonian systems involving elastoviscoplastic fluids~\cite{oishi_thompson_martins_2019} would be interesting.  

\section*{Acknowledgements}

The first author would like to thank the financial support given by Sao Paulo Research Foundation (FAPESP) grants \#2013/07375-0  and \#2021/13833-7, and the National Council for Scientific and Technological Development (CNPq), grant \#305383/2019-1. The authors acknowledge support from the National Science Foundation AI Institute in Dynamic Systems
(grant number 2112085) and from the Army Research Office (ARO W911NF-19-1-0045). The authors also thank Paolo Conti and Samuel E. Otto for all their fruitful suggestions and Hugo L. Franca for having shared his Basilisk's script. 

\section*{{Appendix: Results for high Weissenberg number}}

{In this section, we investigate the capability of SINDy ROMs to predict the dynamics of a more complex flow. From a physical point-of-view, as mentioned in the introduction, increasing the Weissenberg number leads to high-oscillatory regime, and the flow presents more chaotic behaviors (see Fig. \ref{oscill_chaotic}b)).}

\begin{figure}[t!]
\vspace{-.1in}
    \centering
\includegraphics[width=1\textwidth]{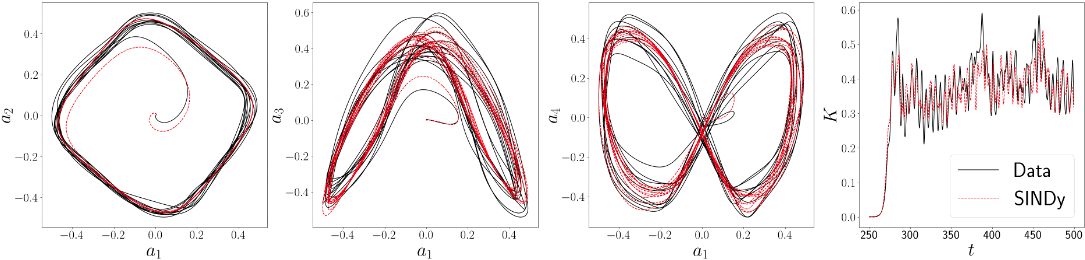} 
\vspace{-.2in}
   \caption{Comparison of the mode pair trajectories in time considering the pairs $(a_1,a_2)$, $(a_1,a_3)$ and $(a_1, a_4)$ and the quantity $K = \frac{1}{2}\boldsymbol{a} \cdot \boldsymbol{a}$ value using the full numerical solution (Data) and the reduced-order model (SINDy) for four-roll mill geometry for $Wi=5.5$.}
    \label{4RcompWi5.5}
\end{figure}

\begin{figure}[t!]
    \centering
\includegraphics[width=0.8\textwidth]{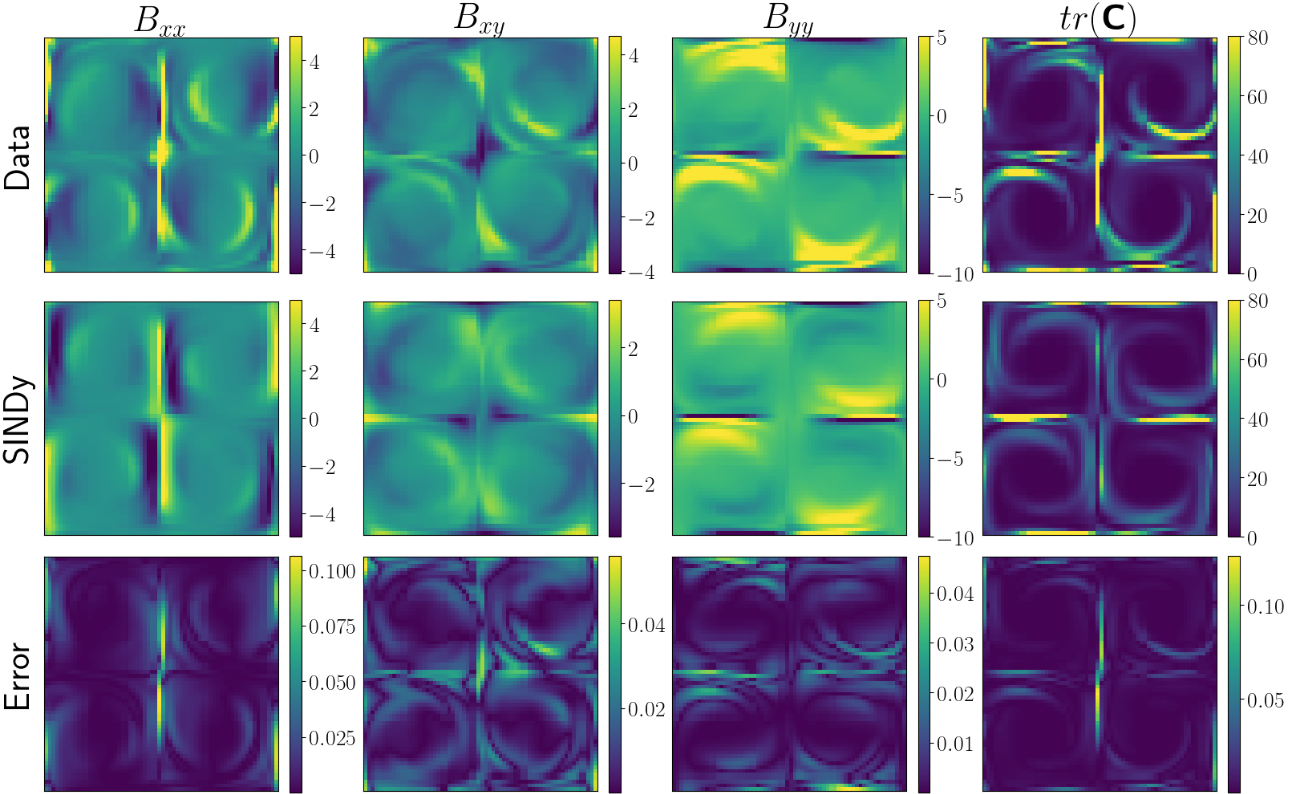}
\vspace{-.1in}
\caption{Comparison between the  full numerical solution (Data), the reduced-order model (SINDy), and the error for $n=1$ and $Wi=5.5$ at simulation time $t=280$.}  
    \label{recon_Wi5.5_1999}
\end{figure}

{In order to extend the framework presented in Section \ref{SecR1} for a fixed high Weissenberg number, we have selected the value $Wi=5.5$ to obtain a learned SINDy model when the flow is aperiodic. Using linear and cubic functions in order to preserve the stability of the model, the identified system is now given by
\begin{equation}
\begin{split}
\begin{pmatrix}
        \dot{a}_{1}   \\
        \dot{a}_{2} 
    \end{pmatrix} 
&    = \begin{pmatrix}
        0.02 & -0.27 \\
        0.34 & 0.01
    \end{pmatrix}
    \begin{pmatrix}
        {a}_{1}   \\
        {a}_{2} 
    \end{pmatrix} 
\\
&+
\begin{pmatrix}
          -0.06{a}_{1}^{2}-0.21{a}_{2}^{2}-0.13{a}_{1}{a}_{2}  & 0.02{a}_{2}^{2} \\
        -0.06{a}_{1}^{2} & -0.19{a}_{1}^{2}-0.04{a}_{2}^{2}+0.08{a}_{1}{a}_{2}
    \end{pmatrix}
    \begin{pmatrix}
        {a}_{1}   \\
        {a}_{2}   \\
    \end{pmatrix}. 
    \label{si3}
\end{split}
\end{equation}
Comparing with system (\ref{si}), the resulting system (\ref{si3}) breaks symmetry in the model coefficients. We used relation (\ref{rel1}) to construct the third and fourth modes as functions of the first two modes. The trajectories of the mode pairs are depicted in Fig. \ref{4RcompWi5.5}, which also includes the quantity $K$. It is important highlight, as shown in this figure, model (\ref{si3}) qualitatively captures the transitions observed in this aperiodic flow for all modes employed in the ROM, maintaining stability even at a high Weissenberg number. Additionally, the reconstruction errors, as detailed in Fig. \ref{recon_Wi5.5_1999}, demonstrate a reseable agreement between the data and the SINDy-learned model, with the maximum error reaching approximately $10\%$. This provides further evidence that our framework has the potential to capture more intricate dynamics inherent in the four-roll geometry, such as elastic turbulence.}

\begin{figure}[t!]
    \centering
   \includegraphics[scale=0.5]{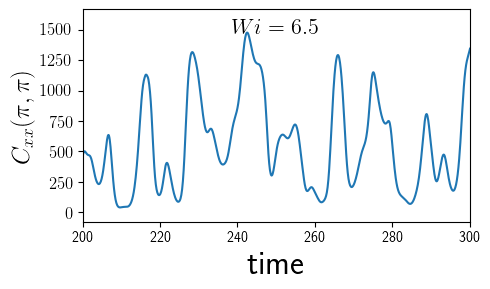} \quad 
      \includegraphics[scale=0.5]{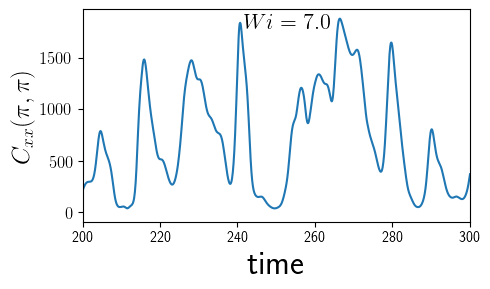}
    \caption{Time series of the $C_{xx}$-component of the conformation tensor at the central stagnation point of the four-roll mill geometry with $n=1$.}
    \label{chaotic}
\end{figure}

\begin{figure}[t!]
	\centering
	\includegraphics[scale=0.5]{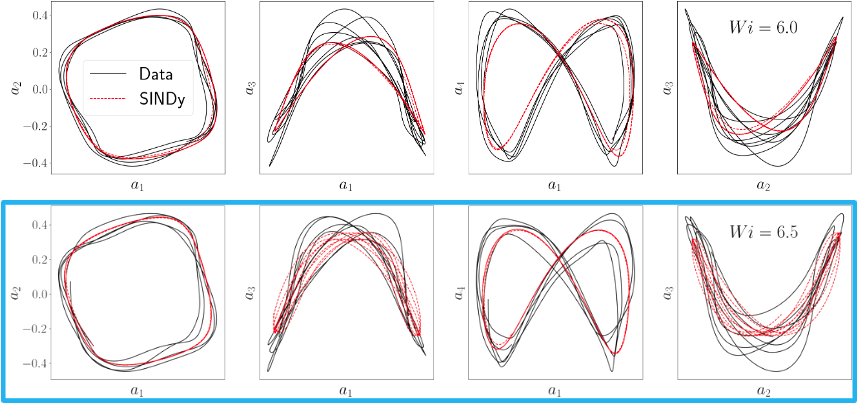}\\
	\includegraphics[scale=0.5]{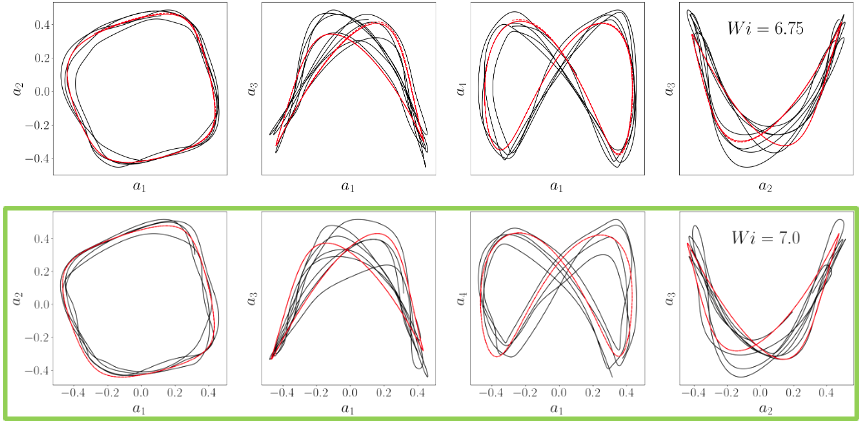}\\
	\caption{{Comparison of the phase portraits for the pairs $(a_1,a_2)$, $(a_1,a_3)$, $(a_1, a_4)$ and $(a_2, a_3)$ using the full numerical solution (Data) and the reduced-order model (SINDy) for $n=1$ with training data $Wi=\{6,6.75\}$, with an interpolated value of $Wi=6.5$ (highlighted in blue) and an extrapolated value of $Wi=7$ (highlighted in green) for testing data.}}
	\label{4Rcomp3-2Wi6}
	\vspace{-.2in}
\end{figure}

{As a further investigation, we applied the method described in Section 5 to identify a parameterized system with higher values of $Wi$. The temporal window is $t \in [200, 300]$, resulting in $N=1000$ snapshots for training values $Wi={6,6.75}$, and testing on the values $Wi=6.5$ and $Wi=7.0$. Both the interpolated and extrapolated test values exhibit chaotic dynamics, as shown in Fig. \ref{chaotic}. The parameterized model maintains its stability even for these higher values of $Wi$, and the predicted solutions qualitatively match the true solutions, as can be seen in Fig. \ref{4Rcomp3-2Wi6}. To enhance the accuracy of the ROM in capturing the solutions of more chaotic dynamics, there is a need to include more modes and data into the SINDy algorithm, which will be the focus of ongoing work. However, even in these cases the reconstruction errors remain below 10$\%$, showcasing the framework's potential for further enhancements.}

\clearpage
\begin{spacing}{.84}
\small
\bibliographystyle{unsrt}

\end{spacing}
\end{document}